\documentclass[preprint,superscriptaddress]{revtex4-1}

\usepackage{graphicx}
\usepackage{amsmath}
\usepackage{amssymb}
\usepackage{color}
\usepackage[colorlinks,linkcolor=blue,
            citecolor=blue,
            hyperindex,
            pdfstartview=FitH,
            plainpages=false]
            {hyperref}

\newcommand {\bisco}{Bi$_2$Sr$_2$CaCu$_2$O$_{8+\delta}$}
\newcommand {\uJcm}{$\mu$Jcm$^{-2}$}

\begin{document}
\title{Ultrafast quenching of electron-boson interaction and superconducting gap in a cuprate superconductor}
\author{Wentao Zhang}
\affiliation{Materials Sciences Division, Lawrence Berkeley National Laboratory, Berkeley, California 94720, USA}
\affiliation{Department of Physics, University of California, Berkeley, California 94720, USA}
\author{Choongyu Hwang}
\affiliation{Materials Sciences Division, Lawrence Berkeley National Laboratory, Berkeley, California 94720, USA}
\affiliation{Department of Physics, Pusan National University, Busan 609-735, Republic of Korea}
\author{Christopher L. Smallwood}
\affiliation{Materials Sciences Division, Lawrence Berkeley National Laboratory, Berkeley, California 94720, USA}
\affiliation{Department of Physics, University of California, Berkeley, California 94720, USA}
\author{Tristan L. Miller}
\affiliation{Materials Sciences Division, Lawrence Berkeley National Laboratory, Berkeley, California 94720, USA}
\affiliation{Department of Physics, University of California, Berkeley, California 94720, USA}
\author{Gregory Affeldt}
\affiliation{Materials Sciences Division, Lawrence Berkeley National Laboratory, Berkeley, California 94720, USA}
\affiliation{Department of Physics, University of California, Berkeley, California 94720, USA}
\author{Koshi Kurashima}
\affiliation{Department of Applied Physics,Tohoku University, Sendai 980-8579, Japan}
\author{Chris Jozwiak}
\affiliation{Advanced Light Source, Lawrence Berkeley National Laboratory, Berkeley, California 94720, USA}
\author{Hiroshi Eisaki}
\affiliation{Electronics and Photonics Research Institute, National Institute of Advanced Industrial Science and Technology, Ibaraki 305-8568, Japan}
\author{Tadashi Adachi}
\affiliation{Department of Engineering and Applied Sciences, Sophia University, Tokyo 102-8554, Japan}
\affiliation{Department of Applied Physics,Tohoku University, Sendai 980-8579, Japan}
\author{Yoji Koike}
\affiliation{Department of Applied Physics,Tohoku University, Sendai 980-8579, Japan}
\author{Dung-Hai Lee}
\affiliation{Department of Physics, University of California, Berkeley, California 94720, USA}
\author{Alessandra Lanzara}
\email{alanzara@lbl.gov}
\affiliation{Materials Sciences Division, Lawrence Berkeley National Laboratory, Berkeley, California 94720, USA}
\affiliation{Department of Physics, University of California, Berkeley, California 94720, USA}
\date {\today}

\begin{abstract}
Ultrafast spectroscopy is an emerging technique with great promise in the study of quantum materials, as it makes it possible to track similarities and correlations that are not evident near equilibrium.
Thus far, however, the way in which these processes modify the electron self-energy---a fundamental quantity describing many-body interactions in a material---has been little discussed. 
Here we use time- and angle-resolved photoemission to directly measure the self-energy's ultrafast response to near-infrared photoexcitation in high-temperature cuprate superconductor.
Below the superconductor's critical temperature, ultrafast excitations trigger a synchronous decrease of electron self-energy and superconducting gap, culminating in a saturation in the weakening of electron-boson coupling when the superconducting gap is fully quenched.
In contrast, electron-boson coupling is unresponsive to ultrafast excitations above the superconductor's critical temperature and in the metallic state of a related material.
These findings open a new pathway for studying transient self-energy and correlation effects in solids.
\end{abstract}

\maketitle
The dynamics of electrons and atoms interacting with intense, and ultrashort optical pulses is one of the emerging fields in physics.
Strong optical pulses have been used as powerful tools to measure the electron-phonon interaction in solids\cite{Allen1987,Brorson1990}, to investigate fundamental dynamical processes in semiconductors\cite{GUPTA1992,Ulbricht2011}, and to modulate the lattice structure of solids by creating dynamical states with new properties\cite{Schmitt2008,Tomeljak2009,Rohwer2011,Mizoguchi2013}.
These methods are particularly exciting in the context of correlated materials, where intense optical fields can drive a transition from an insulating to a metastable metallic phase\cite{Rini2005}, can induce transient signatures of superconductivity\cite{Hu2014}, can lead to anisotropic modulation of the electron-phonon coupling\cite{Carbone2008}, and can disentangle the different dynamics in governing the superconducting and pseudogap phase of cuprates\cite{Demsar1999,Kaindl2000,Hinton2013,Coslovich2013}.

Despite the large amount of new physics revealed, most studies thus far use all-optical techniques, which do not directly probe quasiparticles or carry any momentum information.
As a consequence, the way fundamental quantities such as the electron self-energy and many-body interactions evolve outside equilibrium is often inferred indirectly.
When probed in a time-resolved manner these quantities have the potential to reveal insights on the microscopic properties of solids\cite{Allen1987,Sentef2013}.
Recent developments in high-resolution time- and angle-resolved photoemission spectroscopy (trARPES) now make these studies possible.
So far, however, most of the trARPES studies have focused on recombination dynamics of photo-induced quasiparticle population\cite{Perfetti2007,Graf2011,Corts2011,Smallwood2012,Zhang2013a} and gap dynamics\cite{Schmitt2008,Rohwer2011,Brouet2013,Smallwood2014}.

Here we present a study on the high-temperature cuprate superconductor \bisco\ (Bi2212), and compare it to metallic Bi$_{1.76}$Pb$_{0.35}$Sr$_{1.89}$CuO$_{6+\delta}$ (Bi2201).
In cuprate materials like these, there is known to be a universal electron self-energy renormalization effect (a kink in the dispersion), which signifies the coupling of the electrons to bosons\cite{Bogdanov2000,Lanzara2001,Kaminski2001,Johnson2001,Gweon2004}.
However, whether this kink is related in any way to superconductivity is highly debated.
Using trARPES, we tracked the temporal evolution of the electron self-energy renormalization and the superconducting gap after the system is perturbed by a femtosecond pump pulse.
We found that, in superconducting Bi2212, the real part of the self-energy and the superconducting gap are dramatically suppressed.
Both effects also saturate at the same pump fluence.
In contrast, in the normal state of Bi2212 and metallic heavily overdoped Bi2201, the suppression of the electron self-energy is far weaker.
The results open a new avenue of investigation into self-energy and electron-boson interactions in solids.

Data were taken on an ultra-high resolution setup as previously described\cite{Smallwood2012a}.
In order to avoid the effects of photon-induced electric fields\cite{Sentef2013}, we limit our study to relatively low fluence (between 4 and 24 \uJcm\ or between 0.004 V/$a_0$ and 0.009 V/$a_0$) and long time delay ($\geq300$ fs).
However, the fluence used in this study is still high enough to drive the full closure of the superconducting gap on the Fermi arc\cite{Smallwood2014}. 

\section*{Results}
\begin{figure}
\centering\includegraphics{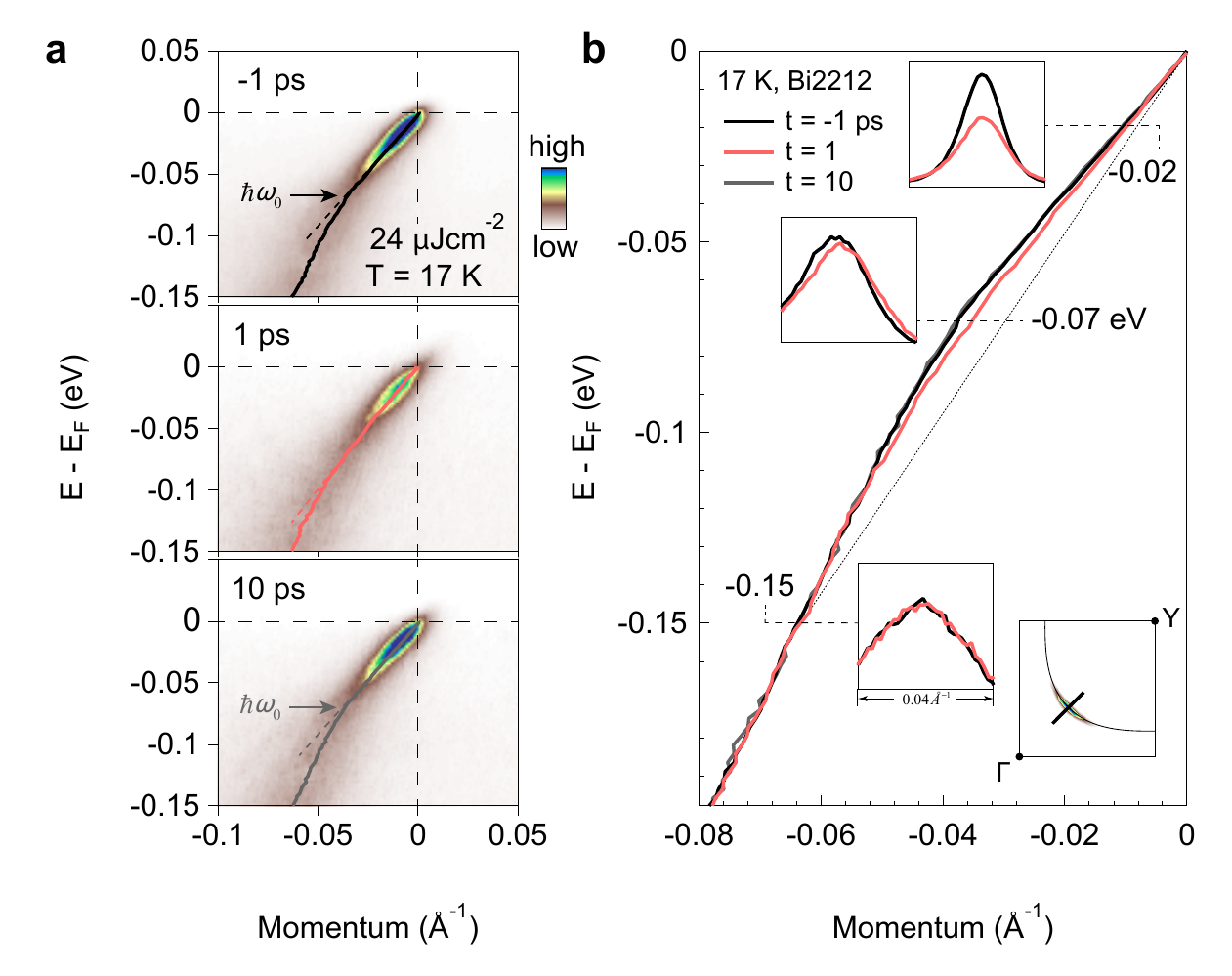}
  \caption{
Time-resolved spectra on a nearly optimally doped Bi2212.
$\textbf{a}$, Equilibrium (before pumping, $t=-1$ ps) and transient (after pumping, $t = 1$ ps and $t = 10$ ps) photoelectron intensity (represented by false color) as a function of energy and momentum  measured along a nodal cut, for a pump fluence of 24 \uJcm. The bold solid black lines are the momentum distribution curve (MDC) dispersions at the corresponding delay time. The arrows mark the position of the kink at $\hbar\omega_0$ $\sim$70 meV.
$\textbf{b}$, MDC dispersions for different delay times ($-1$, $1$, and $10$ ps).  Insets show comparisons of MDCs before pumping ($-1$ ps) and after pumping (1 ps) for a series of binding energies ($-0.15$, $-0.07$, and $-0.02$ eV).
}
\label{Fig1}
\end{figure}

\textbf{Electronic dispersion.}
Figure~\ref{Fig1}a shows the equilibrium ($t=-1$ ps), and non-equilibrium ($t=1$ and $t=10$ ps) ARPES intensity as a function of energy and momentum for a nearly optimally doped sample measured with pump fluence 24 \uJcm~at 17 K, far below $T_\text{c}$, along the nodal direction.
The equilibrium spectrum shows the widely studied renormalization kink at $\sim$70 meV\cite{Bogdanov2000,Lanzara2001,Kaminski2001,Johnson2001}.
Upon pumping, at a delay time of 1 ps, a clear loss of spectral weight can be observed, which is mainly confined between the Fermi energy and the kink energy\cite{Graf2011}.
It takes approximately 10 ps for the transient spectra to recover back to the equilibrium state.
The effect of laser pumping on the dispersion is shown in Fig. \ref{Fig1}b, where the equilibrium (black solid line) and transient (red solid line) dispersions are compared.
The dispersions are extracted in the standard way by fitting momentum distribution curves (MDCs) to a Lorentzian functional form\cite{Valla1999a}.
The comparison between the dispersion curve after a long delay time ($t = 10$ ps) (gray line) with the equilibrium curve provides an estimate of our error bars. 
The most obvious pump-induced change occurs near the kink energy, as shown by the shift of MDC peak position in this energy range, in contrast to the MDCs near the Fermi level or at much higher binding energies, where the shift is negligible (see inset).
Specifically, for delay time $t=1$ ps, the Fermi velocity increases by 0.13 eV$\cdot$\AA (equilibrium 1.87 eV$\cdot$\AA ) at binding energy below the kink energy and remains approximately unchanged above, resulting in an apparent softening of the kink strength (such as coupling strength) as pumping is turned on.
The Fermi velocity is extracted from the slope in the dispersion between $\sim$70 meV and $E_\text{F}$, as $v = dE/dk$\cite{Ashcroft1976}.

\begin{figure}
\centering\includegraphics{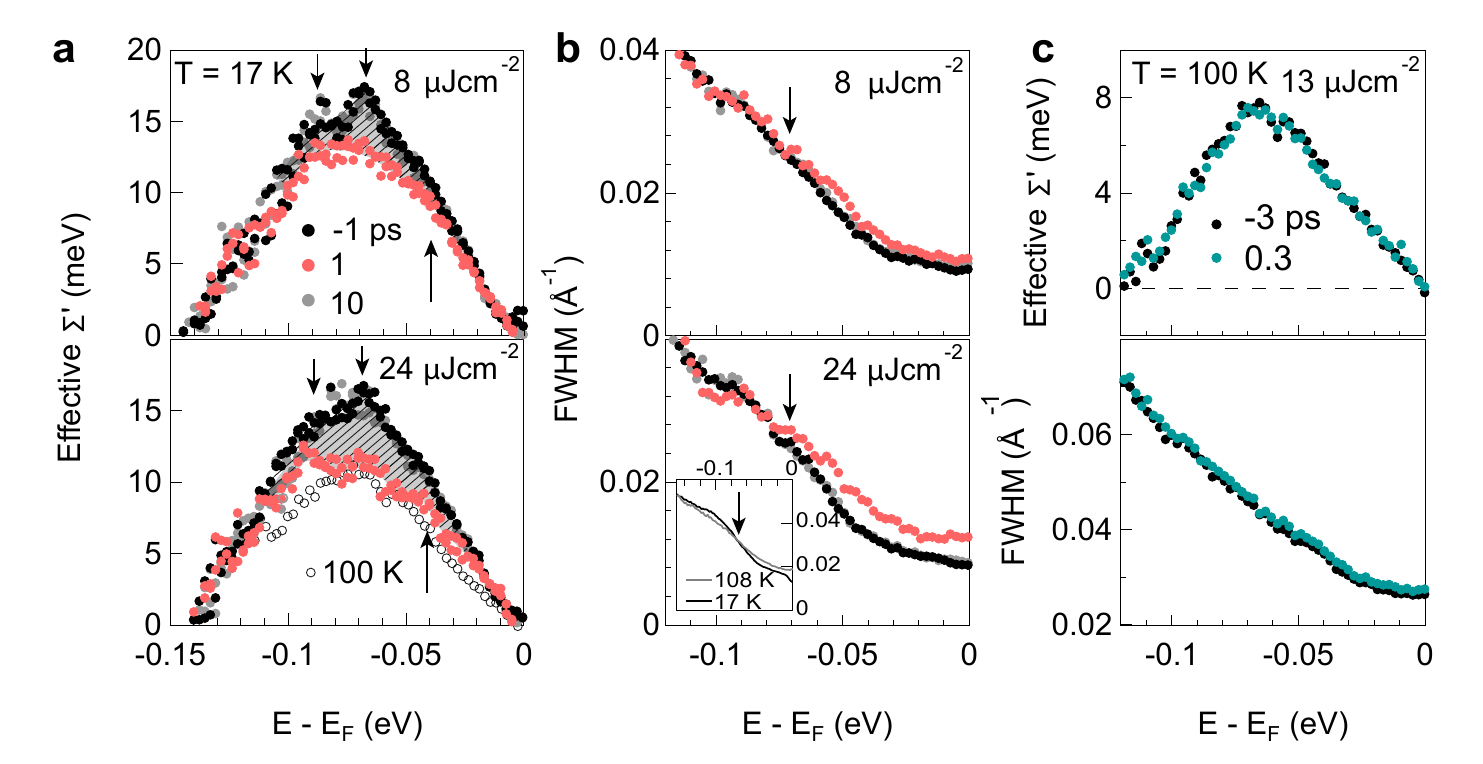}
  \caption{
Equilibrium and transient nodal electron self-energies.
$\textbf{a}$, Real part of electron self-energy ($\Sigma'$) at different delay times ($t = -1$, $1$ and $10$ ps) for pump fluences 8 and 24 \uJcm measured at 17 K (below $T_\text{c}$).  $\Sigma'$ measured at equilibrium temperature 100 K (above $T_\text{c}$) is plotted in the lowest panel for comparison.
$\textbf{b}$, The corresponding MDC width as a function of energies. The black arrows mark the energy $\sim$$\hbar\omega_0$ where non-equilibrium and equilibrium MDC width separate with each other.
Inset in the lowest panel shows non-equilibrium MDC width at 17 K and 108 K\cite{Zhang2008}.
$\textbf{c}$, $\Sigma'$ and MDC width, similar to that shown in $\textbf{a}$ and $\textbf{b}$,
 but measured in the normal state with pump fluence 13 \uJcm, at an equilibrium temperature of 100 K.
}
  \label{Fig2}
\end{figure}

\begin{figure}
\centering\includegraphics{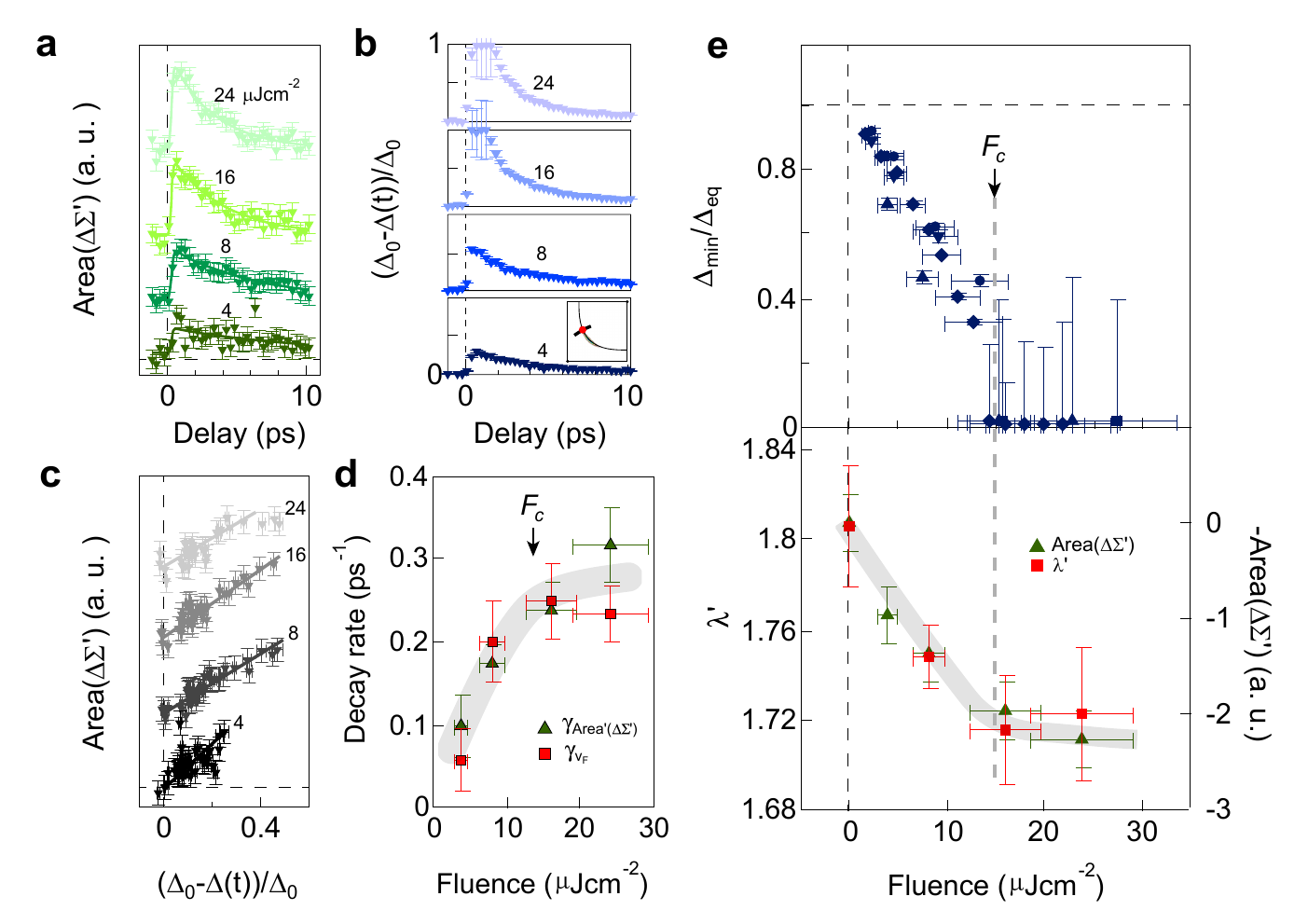}
  \caption{
Dynamics of pump-induced change.
$\textbf{a}$, Pump-induced change of nodal $\Sigma'$ (vertically offset for clarity) as a function of delay time for different fluences. The error bars of Area($\Delta$$\Sigma'$) are estimated by the maximum difference of Area($\Delta$$\Sigma'$) among the measurements in equilibrium state.
$\textbf{b}$, Gap vs pump-probe delay time for an off-nodal cut (inset) at different fluences. The changes of energy gap are normalized to the maximum gap.
$\textbf{c}$, Pump-induced change of the $\Sigma'$ (offset for clarity) as a function of the energy gap at each delay time as extracted from $\textbf{a}$ and $\textbf{b}$, for four different fluences.
Because of the limited energy resolution (23 meV, comparable with the equilibrium gap size), only energy gap changes less than 50\% are reliable and shown.
$\textbf{d}$, The recovery rate of nodal $\Sigma'$ and the nodal coupling strength $\lambda'$ as a function of fluence.
$\textbf{e}$, The pump-induced non-equilibrium minimum energy gap, maximum changes of Area($\Delta$$\Sigma'$) and electron-boson coupling strength at 70 meV as a function of pump fluence.
The coupling strength is given by $\lambda'=v_0/v_\text{F}$ as functions of pump fluence ($v_0$ ($v_\text{F}$), bare (Fermi) velocity below (above) the kink energy)
The error bars of the change of $\Sigma'$ and $\lambda'$ are absolute maximum variations before $t=0$, and the others are standard deviations from fitting.
}
 \label{Fig3}
\end{figure}

\textbf{Self-energy changes below $T_\text{c}$.}
The differences between equilibrium and transient spectra can be analyzed by extracting the electron self-energy $\Sigma=\Sigma'+i\Sigma''$, shown in Fig. \ref{Fig2}.
To extract the effective real part of the self-energy at different delay time, we took measured dispersions and subtracted featureless linear bare bands with the same velocity at each delay time (see, for example, the dotted line in Fig. \ref{Fig1}b).
Such techniques are commonly employed to analyze the electron-boson interaction at equilibrium\cite{Valla1999a,Bogdanov2000}.
At equilibrium (black dots) ${\Sigma}'$ is reminiscent of a spectrum of modes, as previously reported\cite{Kaminski2001,Johnson2001,Zhang2008,Lee2008}, and its maximum is at the 70 meV kink position.
The most significant pump-induced effect is the suppression of $\Sigma'$ in the proximity of the kink energy (between 40 and 90 meV) (Fig. \ref{Fig2}a) (red dots), in line with a softening of the coupling strength, as shown in Fig. \ref{Fig1}.
This suppression intensifies as fluence increases, and eventually saturates when superconductivity is completely suppressed (see Fig. \ref{Fig3}), as we will discuss later.
Within our resolution, we observe no pump-induced energy shift in the peak position of $\Sigma'$\cite{Kulic2005}.
In the same figure (bottom panel of Fig. \ref{Fig2}a) we directly compare the  equilibrium $\Sigma'$ at 100 K with the pump-induced $\Sigma'$ at a similar electronic temperature as measured from the width of the Fermi edge\cite{Perfetti2007,Graf2011,Smallwood2014}.
While the effect of temperature on the equilibrium $\Sigma'$ extends over the entire energy range, the effect of pumping is smaller overall and is mainly confined within the kink energy (see also Supplementary Figure 1).  
This suggests that optical pumping induces an effect beyond increasing the temperature.

In Fig. \ref{Fig2}b we show the imaginary part of the self-energy, $\Sigma''$, which is proportional to the full width at half maximum (FWHM) of the MDCs.
In agreement with $\Sigma'$, the main pump-induced change is confined between the kink energy and the Fermi level (vertical black arrow) and increases as fluence increases.
This is in contrast to thermal effects, where the temperature causes a change of $\Sigma''$ over the entire energy window (see inset of panel b and Supplementary Figure 1).
In Fig. \ref{Fig2}c we show the comparison of equilibrium and transient $\Sigma'$ and $\Sigma''$ at equilibrium temperature 100 K, above $T_\text{c}$.
In sharp contrast with the low-temperature behavior, the pump-induced changes of the self-energy are negligible in the normal state up to the highest applied fluence 24 \uJcm~(we note, however, that this might no longer be valid at very high fluences, which would substantially affect the entire band structure\cite{Sentef2013,Rameau2014}.
These results might suggest that the pump-induced changes of self-energy are sensitive to the presence of superconductivity and are not induced by trivial thermal broadening effects.

We note that the absence of a shift in the kink energy, when the superconducting gap closes, suggests that the electron-boson interaction falls beyond the standard Migdal-Eliashberg theory for superconductivity as also suggested by equilibrium ARPES experiments\cite{He2013,Plumb2013}.
Indeed, within this standard theory the energy dispersion has a square-root divergence at energy $\Delta_0+\Omega$ in $\Sigma'(\omega)$ ($\Omega$ is the boson energy and $\Delta_0$ is the superconducting gap at zero temperature)\cite{Scalapino1969}.
Therefore, when the pump drives the superconducting gap to zero, one expects a shift of the kink energy and of $\Sigma'$ toward lower binding energy by the magnitude of the superconducting gap, even for the nodal cut\cite{Sandvik2004}.

\textbf{Self-energy versus superconducting gap.}
To further investigate this matter, we utilize the unique advantage of trARPES by simultaneously monitoring the pump-induced changes in both the electron self-energy and the superconducting gap.
Figure~\ref{Fig3} compares the pump-induced change in $\Sigma'$ and the non-equilibrium superconducting gap.
The latter is measured for a cut on the Fermi arc (see inset of Fig. \ref{Fig3}b).
Energy gaps at each delay time are obtained by fitting symmetrized electron distribution curves to an energy-resolution-convolved phenomenological BCS model\cite{Norman1998}, which is widely used in characterizing the energy gap in cuprates\cite{Lee2007,Smallwood2014} (see also Supplementary Figure 2).   
The temporal evolution of the area of $\Delta \Sigma'$
(hatched area in Fig. \ref{Fig2}a) and the superconducting gap for different fluences are shown in panels a and b of Fig. \ref{Fig2}, respectively.
In agreement with Fig. \ref{Fig2}, $\Sigma'$ is weakly affected at low fluences and shows substantially slower initial recovery rate (0.1 ps$^{-1}$) than at higher fluences (0.32 ps$^{-1}$; Fig. \ref{Fig3}d. 
Similarly, pumping only weakly affects the superconducting gap at low fluence (bottom measurements in Fig. \ref{Fig3}b) and eventually drives it to a full closure at high fluences (top measurements)\cite{Smallwood2012,Smallwood2014}.
A similar fluence dependence is also observed for the Fermi velocity (see Supplementary Figure 3).
The pump-induced change in the self-energy and the non-equilibrium superconducting gap at each delay time are plotted in Fig. \ref{Fig3}c for several fluences, showing an unexpected linear relation for all fluences.
At the highest fluence a small deviation from linearity is observed, possibly because of additional broadening of the spectra, causing an underestimate of the superconducting gap.

In Fig. \ref{Fig3}e we show the fluence dependence of the maximal near-nodal gap shift (Fig. \ref{Fig3}b) and compare it with the area of the maximal self-energy change (from Fig. \ref{Fig3}a) and the coupling constant.
In a simple electron-boson coupling model\cite{Park2008} the coupling strength $\lambda$ is directly related to the bare and dressed Fermi velocities according to $\lambda' \equiv \lambda + 1= v_{\text{F}}/v_0 $.
Alternatively, the change in $\lambda$ can be approximated by the integral of the self-energy change between equilibrium and pumped values (see hatched areas Fig. 2) because near the Fermi energy $\lambda = \lim_{\omega \to 0} \partial \Sigma'/\partial \omega$, namely, $\Sigma'(\omega) \approx \lambda \omega$.
Thus $\int_0^{\omega_c} \Delta \Sigma'(\omega) d\omega \approx \Delta \lambda \omega_c^2/2$.
Both characterization methods are shown (for more details, see also Supplementary Figure 3).
Interestingly, as we drive the superconducting gap to a gradual melting by increasing the excitation density, we find that the electron-boson interaction decreases in a similar fashion and eventually saturates (see also Supplementary Figure 4) when the superconducting gap is fully quenched.
The saturation effect is in contrast with thermal broadening, where continuous smearing of the kink occurs as the temperature gradually increases\cite{Zhang2008}.

Figure \ref{Fig3}d shows the recovery rate of the integrated self-energy change (from Fig. \ref{Fig3}a) and $\lambda'$ (see Supplementary Figure 3).
In the low fluence regime we find that both of these rates increase linearly with fluence, in a similar fashion to the bimolecular recombination of non-equillibrium quasiparticles\cite{Gedik2004,Smallwood2012}.
As the fluence approaches the critical fluence, both rates saturate, marking the onset of different recombination processes, a behavior which appears to be dictated by the closing of the superconducting gap at $F_c$.
The behavior is also consistent with previous reports of non-equilibrium quasiparticle recombination in an optimally doped and underdoped Bi2212\cite{Smallwood2012,Zhang2013a}, and with a time-resolved optical reflectivity study\cite{Coslovich2011a}. 

\begin{figure}
\centering\includegraphics{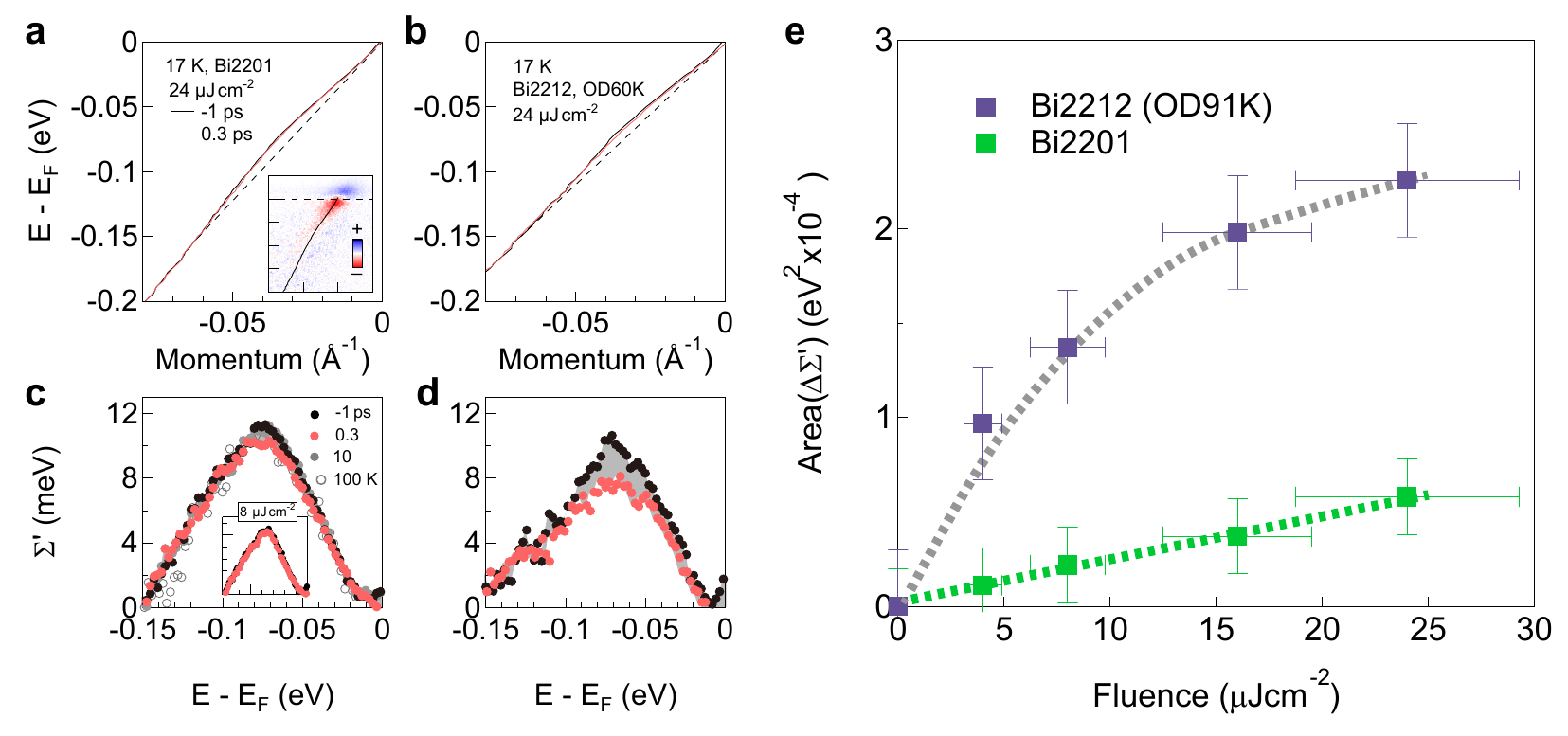}
  \caption{
Results on a heavily overdoped Bi2201 and an overdoped Bi2212.
$\textbf{a}$ and $\textbf{b}$, MDC dispersions of equilibrium ($t = -1$ ps) and non-equilibrium ($t = 0.3$ ps) states, for a pump fluence of 24 \uJcm\ for the overdoped Bi2201 and the overdoped Bi2212, respectively.
False colored inset in $\textbf{a}$ shows the difference between equilibrium ($t=-1$ ps) and non-equilibrium ($t=0.3$ ps) raw spectral image.
$\textbf{c}$ and $\textbf{d}$, Real part of the electron self-energy ($\Sigma'$) at different delay time ($t=-1$, 0.3 and 10 ps) for pump fluence 24 \uJcm\ for the overdoped Bi2201 and the overdoped Bi2212, respectively.
The equilibrium $\Sigma'$ measured at 100 K on the overdoped Bi2201 is represented by empty circles for comparison in $\textbf{c}$.
The inset in $\textbf{c}$ shows $\Sigma'$ spectra for a pump fluence of 8 \uJcm.
$\textbf{e}$, Fluence dependence of the Area($\Sigma'$) for the optimally doped Bi2212 and heavily overdoped Bi2201. The bold lines are guides to the eye.
The error bars in \textbf{d} are absolute maximum variations before $t=0$.
}
\label{Fig4}
\end{figure}

\textbf{Self-energy changes above $T_\text{c}$.} 
To better understand the nature of the reported pump-induced change in the self-energy of Bi2212 superconductor, in Fig. \ref{Fig4} we study the effect of pumping on the electron-boson coupling in a metal in the same low fluence regime and at the same low temperature. 
To make a meaningful comparison with Bi2212, we report data for an heavily overdoped Bi2201 ($T_c<2$ K).
Bi2201 has very similar crystallographic and electronic structure to Bi2212, and an only slightly weaker 70 meV dispersion kink\cite{Lanzara2001}.
Moreover, Bi2201 can be grown in the heavily overdoped regime, with $T_\text{c}$ lower than a few Kelvin, making it possible for us to study its normal state at 17 K, the same temperature as the study of optimally doped Bi2212.

Figure~\ref{Fig4}a shows the pump-induced changes in the electronic dispersion of heavily overdoped metallic Bi2201 along the nodal direction. 
Although qualitatively the pump-induced changes of spectral intensity are similar to those in superconducting optimally doped Bi2212\cite{Graf2011}, there are significant differences between the two. 
Despite the apparent kink feature, the pump-induced changes in $\Sigma'$ are negligible at lower fluences (inset of Fig. \ref{Fig4}c), and show a very small increase with fluence over the entire range.
Note that, consistent with the notion that the critical fluence marks the threshold above which all Cooper pairs are broken\cite{Zhang2013a}, no critical fluence is observed in the $\Sigma'$ in the normal state of this heavily overdoped sample (Fig. \ref{Fig4}e).
This difference is not likely due to the weaker kink in Bi2201.
Indeed, at the highest fluence (Fig. \ref{Fig4}c) the change of $\Sigma'$ is still at least a factor of 2 smaller than the corresponding change observed in overdoped Bi2212---displayed in Fig. \ref{Fig4}d---despite the comparable  kink strength.
The change of $\Sigma'$ in Bi2201 is a factor of 4 smaller than optimally doped Bi2212, where the kink strength is larger only by a factor of 40\% (see Fig. \ref{Fig4}e and the coupling strength in Supplementary Figure 5).
In contrast to Bi2212, the pump-induced changes in Bi2201 can be accounted for by thermal smearing of the dispersion with temperature\cite{Zhao2011,Lee2008}, as shown by the comparison of  equilibrium and transient self-energy at an equilibrium temperature of 100 K.
We caution that these considerations apply to the low pump fluence regime.
This behavior is also consistent with the expected behavior for a metal where temperature smears the kink feature and reduces the magnitude of the maximum of $\Sigma'$ without any change in the electron-boson coupling\cite{Scalapino1969}.

\section*{Discussion}

The effects observed in this paper cannot be simply explained by transient heating.
Indeed, a comparison at similar temperatures between the temperature dependence of the transient self-energy and the equilibrium self-energy shows quite different trends. At equilibrium the temperature-dependent self-energy changes over a broad range of energies\cite{Zhang2008}, in contrast to the transient self-energy, where the changes are confined to the vicinity of the kink energy. Additionally, the equilibrium self-energy gradually changes as the equilibrium temperature increases, and the effect persists even above $T_\text{c}$. In contrast, the pump-induced change in the self-energy saturates at the critical fluence and doesn¡¯t show any change above $T_\text{c}$. Moreover, the temperature dependence of the density of the initial non-equilibrium quasiparticles population is not proportional to the inverse of the specific heat\cite{Zhang2013a}, as expected in the case of thermal dynamics. Finally, the observation that bi-molecular recombination dominates quasiparticle scattering processes in the low-fluence regime further supports the non-thermal nature of the reported effects\cite{Gedik2004,Smallwood2012}.

These observations, together with the negligible pump-induced effect in the normal state and in a metallic compound (Fig. 4) validate the intrinsic nature of the pump-induced phenomena here reported.
Whether these effects are due to a change of the electron-boson coupling constant as in an equilibrium picture or are a consequence of photo-induced delocalization of charge which leads to a reduction of the Jahn Teller effect\cite{Bianconi1996,Zhao1997} and to an apparent softening of the electron-boson interaction, they suggest the exciting possibility that the effects responsible for the self-energy renormalization are an important booster of a large pairing gap\cite{Kaminski2001,Johnson2001,Gweon2004,Valla1999a}.
trARPES experiments where one can coherently pump a specific phonon mode may provide fundamental new insights into this piece of the puzzle.

\section*{Methods}
\textbf{Time-resolved ARPES.}
In our time-resolved ARPES experiments\cite{Graf2011,Smallwood2012,Smallwood2012a}, an infrared pump laser pulse ($h \nu = 1.48$ eV) drives the sample into a non-equilibrium state. Subsequently an ultraviolet probe laser pulse (5.93 eV) photoemits electrons which are captured by a hemispherical analyzer in an ARPES setup.
The system is equipped with a Phoibos 150 mm hemispherical electron energy analyzer (SPECS).
The spot size (FWHM) of pump and probe beams are $\sim$100 $\mu$m and $\sim$50 $\mu$m, respectively, and the pump fluence is calculated using the same method as previously reported\cite{Smallwood2012}.
The repetition rate of the pump and probe beams used in the measurements is 543 kHz.
Time resolution is achieved by varying the delay time ($t$) between the probe and pump pulses.
For $t<0$ the probe pulse arrives before the pump pulse, and hence corresponds to an equilibrium measurement. For $t>0$ the probe pulse arrives after the pump pulse, and hence corresponds to a non-equilibrium measurement. 
The zero delay and time resolution are determined by a cross correlation of pump and probe pulses measured on hot electrons of polycrystalline gold integrated in a 0.4 eV kinetic energy window centered at 1.0 eV above the Fermi energy.
The total energy resolution is $\sim$22 meV, time resolution is $\sim$310 fs (see Supplementary Figure 6), and the momentum resolution is $\sim$0.003 \AA$^{-1}$ at nodal point at the Fermi energy.
In the electron self-energy analysis, the Lucy-Richardson iterative deconvolution method is applied to mitigate the effect of the energy resolution\cite{lucy1974}.

\textbf{Samples.}
Nearly optimally doped \bisco~($T_c=91$ K) and heavily overdoped Bi$_{1.76}$Pb$_{0.35}$Sr$_{1.89}$CuO$_{6+\delta}$ single crystals were grown by the traveling solvent floating zone method.
The overdoped Bi2212 sample ($T_c=$ 60 K) and heavily overdoped Bi2201 sample ($T_\text{c}<2$ K) were obtained by annealing the as-grown optimally doped Bi2212 and Bi2201 in high pressure oxygen.
The $T_\text{c}$ values of the samples were characterized by superconducting quantum interference measurements. Samples were cleaved \emph{in situ} in ultra-high vacuum with a base pressure less than $5\times10^{-11}$ Torr.


\begin{acknowledgments}
This work was supported by Berkeley Lab’s programs on "Quantum Materials" and 
"Ultrafast Materials" funded by the US Department of Energy, Office of Science, Office of Basic Energy Sciences, Materials Sciences and Engineering Division, under Contract No.\ DE-AC02-05CH11231.
\end{acknowledgments}
\section*{Author contributions}
W.T.Z., C.L.S., and T.M. carried out the measurements on Bi2212.
C.H., C.L.S., and W.T.Z. took the measurement on Bi2201.
W.T.Z. C.H., C.L.S., and T.M. were responsible for data analysis.
Bi2212 samples were prepared by H.E.
Bi2201 sample was prepared by K.K., T.A., and Y.K.
A.L. was responsible for the experimental planning and infrastructure.
All authors contributed to the interpretation and writing of the manuscript.

\section*{Additional information}
The authors declare no competing financial interests.

\newpage

\end{document}


\textbf{SUPPLEMENTARY INFORMATION for}
\title{Ultrafast quenching of electron-boson interaction and superconducting gap in a cuprate superconductor}
\author{Wentao Zhang}
\affiliation{Materials Sciences Division, Lawrence Berkeley National Laboratory, Berkeley, California 94720, USA}
\affiliation{Department of Physics, University of Californi\textbf{a}, Berkeley, California 94720, USA}
\author{Choongyu Hwang}
\affiliation{Materials Sciences Division, Lawrence Berkeley National Laboratory, Berkeley, California 94720, USA}
\affiliation{Department of Physics, Pusan National University, Busan 609-735, Republic of Korea}
\author{Christopher L. Smallwood}
\affiliation{Materials Sciences Division, Lawrence Berkeley National Laboratory, Berkeley, California 94720, USA}
\affiliation{Department of Physics, University of Californi\textbf{a}, Berkeley, California 94720, USA}
\author{Tristan L. Miller}
\affiliation{Materials Sciences Division, Lawrence Berkeley National Laboratory, Berkeley, California 94720, USA}
\affiliation{Department of Physics, University of Californi\textbf{a}, Berkeley, California 94720, USA}
\author{Gregory Affeldt}
\affiliation{Materials Sciences Division, Lawrence Berkeley National Laboratory, Berkeley, California 94720, USA}
\affiliation{Department of Physics, University of Californi\textbf{a}, Berkeley, California 94720, USA}
\author{Koshi Kurashima}
\affiliation{Department of Applied Physics,Tohoku University, Sendai 980-8579, Japan}
\author{Chris Jozwiak}
\affiliation{Advanced Light Source, Lawrence Berkeley National Laboratory, Berkeley, California 94720, USA}
\author{Hiroshi Eisaki}
\affiliation{Electronics and Photonics Research Institute, National Institute of Advanced Industrial Science and Technology, Ibaraki 305-8568, Japan}
\author{Tadashi Adachi}
\affiliation{Department of Engineering and Applied Sciences, Sophia University, Tokyo 102-8554, Japan}
\affiliation{Department of Applied Physics,Tohoku University, Sendai 980-8579, Japan}
\author{Yoji Koike}
\affiliation{Department of Applied Physics,Tohoku University, Sendai 980-8579, Japan}
\author{Dung-Hai Lee}
\affiliation{Department of Physics, University of Californi\textbf{a}, Berkeley, California 94720, USA}
\author{Alessandra Lanzara}
\email{alanzara@lbl.gov}
\affiliation{Materials Sciences Division, Lawrence Berkeley National Laboratory, Berkeley, California 94720, USA}
\affiliation{Department of Physics, University of Californi\textbf{a}, Berkeley, California 94720, USA}
\date {\today}

\maketitle

\renewcommand{\bibnumfmt}[1]{[S#1]}
\renewcommand{\citenumfont}[1]{S#1}
\makeatletter
\makeatletter \renewcommand{\fnum@figure}
{\figurename~S\thefigure}
\renewcommand{\theequation}{S\arabic{equation}}
\makeatother

\begin{figure}\centering\includegraphics{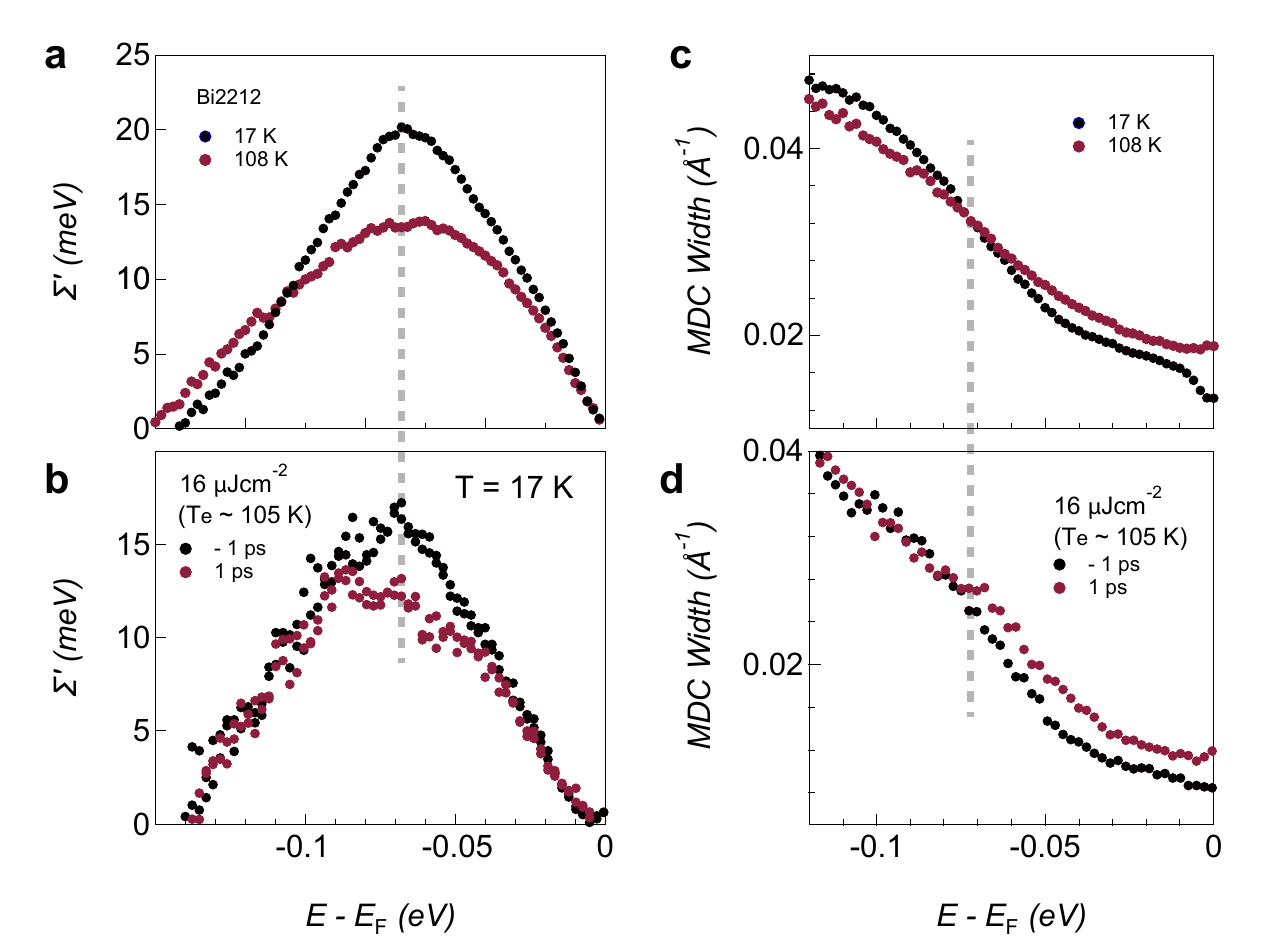}
  \caption{
  Comparison between thermal- and pump-induced change of electron self-energies on Bi2212. \textbf{a}, Equilibrium $\Sigma'$ as a function of energy measured below (17 K) and above (108 K) $T_\text{c}$. \textbf{b}, Non-equilibrium $\Sigma'$ as a function of energy before (t = 1 ps) and after (t = -1 ps) pumping. c and \textbf{d}, similar as in \textbf{a} and  \textbf{b}, but MDC width measured in equilibrium[33] and non-equilibrium states. The pump-probe measurement was taken with pump fluence 16 $\mu$Jcm$^{-2}$.  }
\end{figure}

\begin{figure}\centering\includegraphics[width=1\columnwidth]{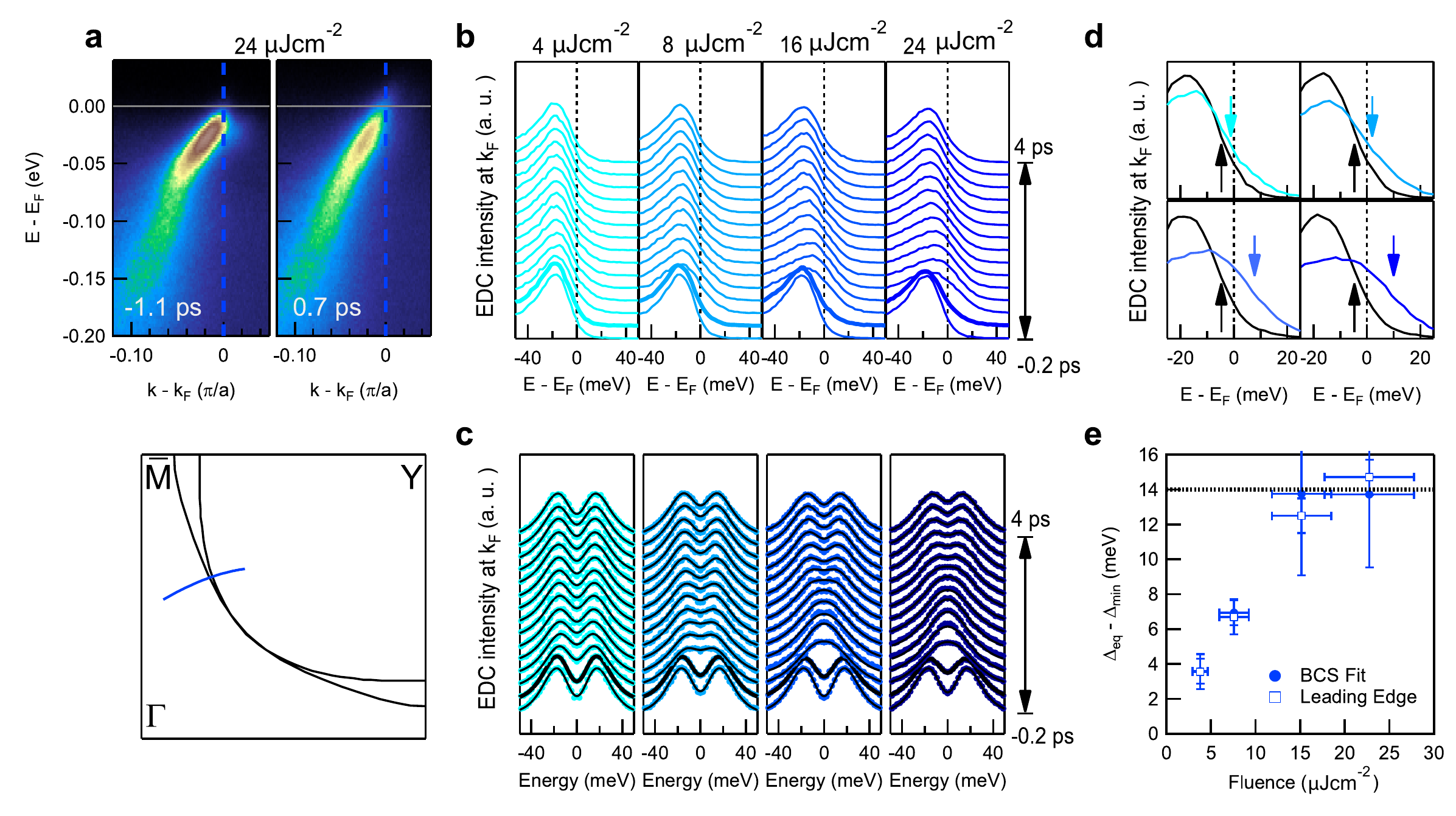}
  \caption{
  Near-nodal gap fitting methodology. \textbf{a}, Off-nodal ARPES dispersion at $\Phi=$ 30$^\circ$ (converted from angle-space into $k$-space and corrected for detector nonlinearity), before and after pumping. \textbf{b}, EDCs at $k_\text{F}$ (intensity along the blue dashed lines in panel a) between -1.1 ps and 4 ps, corresponding to fluences of 4, 8, 16, and 24 $\mu$Jcm$^{-2}$. Bold curves correspond to t = 0 ps. Data are offset vertically for clarity. \textbf{c}, Symmetrized EDCs, acquired by adding the data in a to its reflection across $E_\text{F}$. Black curves correspond to fits using a broadened BCS fit function convolved with a Gaussian resolution function (see Refs. [40] and [41] for details of the functional form). \textbf{d}, Comparison between raw EDCs at equilibrium (shown in black) and those at 0.7 ps (shown in blue) for pump fluences of 4, 8, 16, and 24 $\mu$Jcm$^{-2}$, with colors indexed to the fluence values in panel b. e, Maximal gap shift ($\Delta_\text{eq}-\Delta_\text{min}$) versus pump fluence. Filled circles are the result of the BCS fit. Open squares are from the leading-edge shift in the raw EDCs (see the arrows in panel d), and illustrate the fact that the measurements of the non-equilibrium gap are not unique to the BCS fitting methodology. Extended details about non-equilibrium gap fits can be found in Ref. [41].}
\end{figure}

\begin{figure}\centering\includegraphics{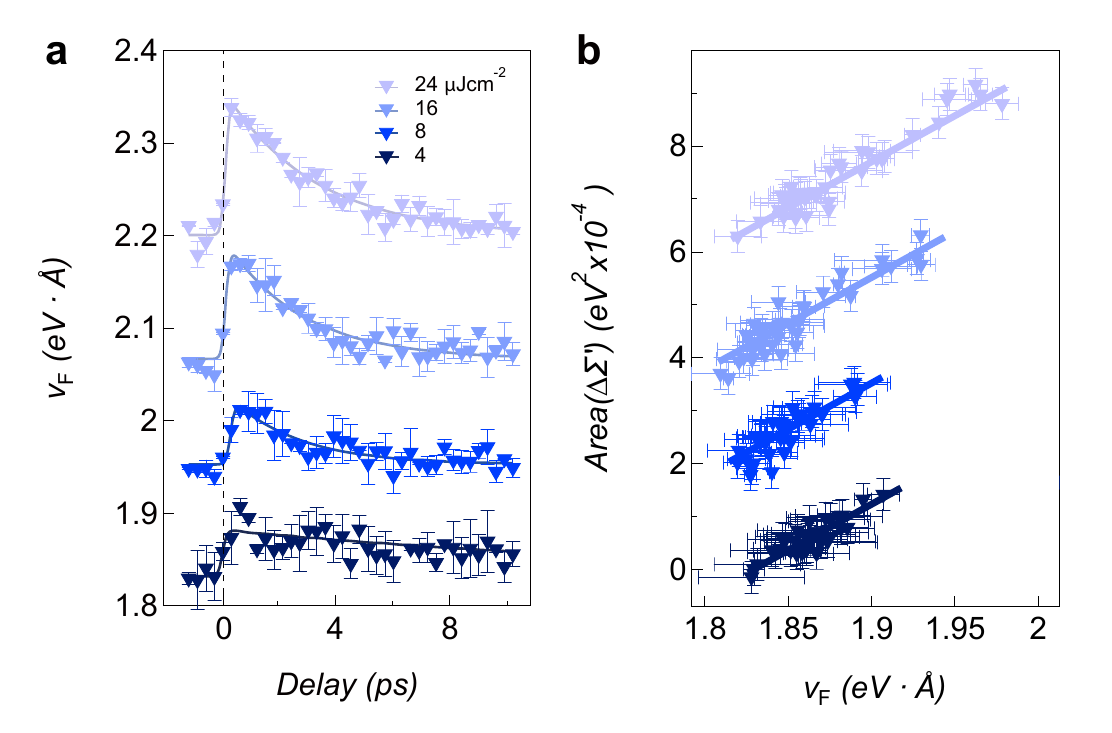}
  \caption{
   Pump-induced weakening of electron-phonon coupling at 17 K. \textbf{a}, Pump-induced suppression of the Fermi velocity. \textbf{b}, Pump-induced change of the real part of the self-energy $\Sigma'$ (offset for clarity) as a function of the Fermi velocity at each delay time.}
\end{figure}

\begin{figure}\centering\includegraphics{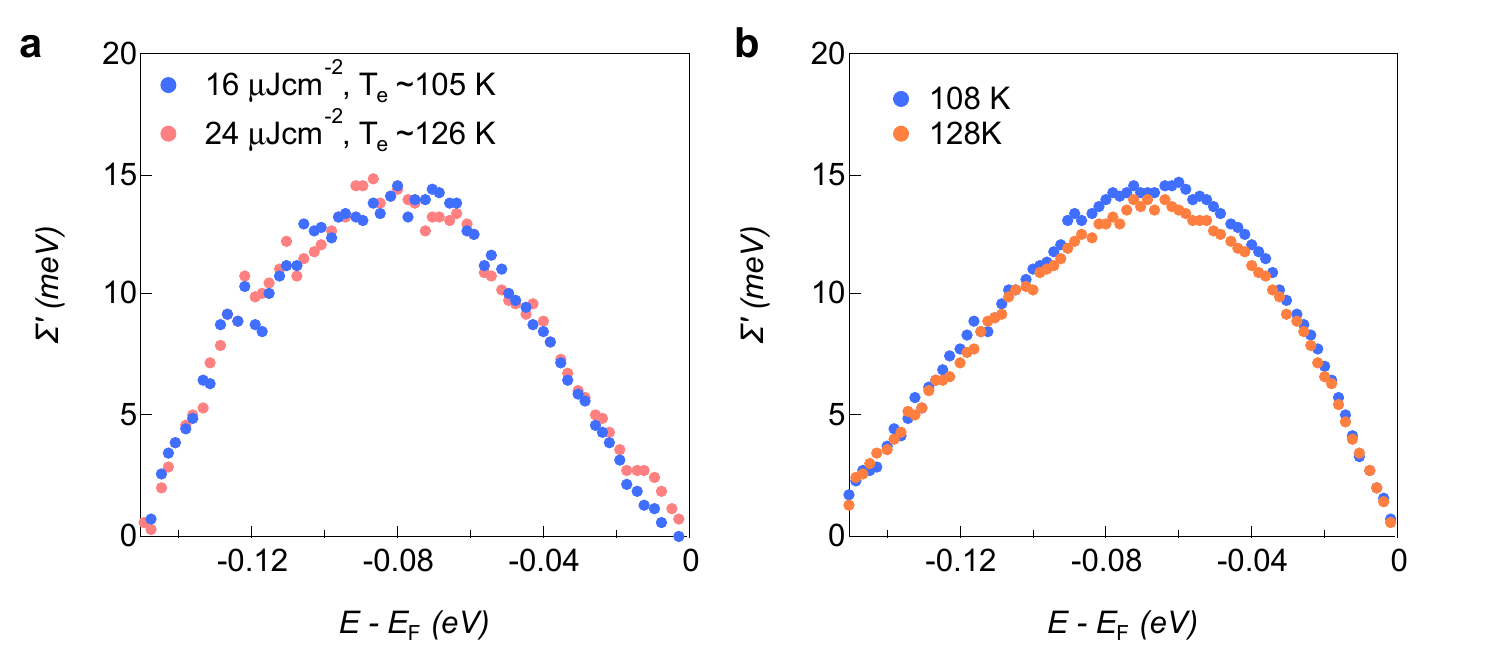}
  \caption{
   Real part of self-energy. Non-equilibrium real part of self-energy for fluences above the $F_\text{c}$ (16 and 24 $\mu$Jcm$^{-2}$). }
\end{figure}

\begin{figure}\centering\includegraphics{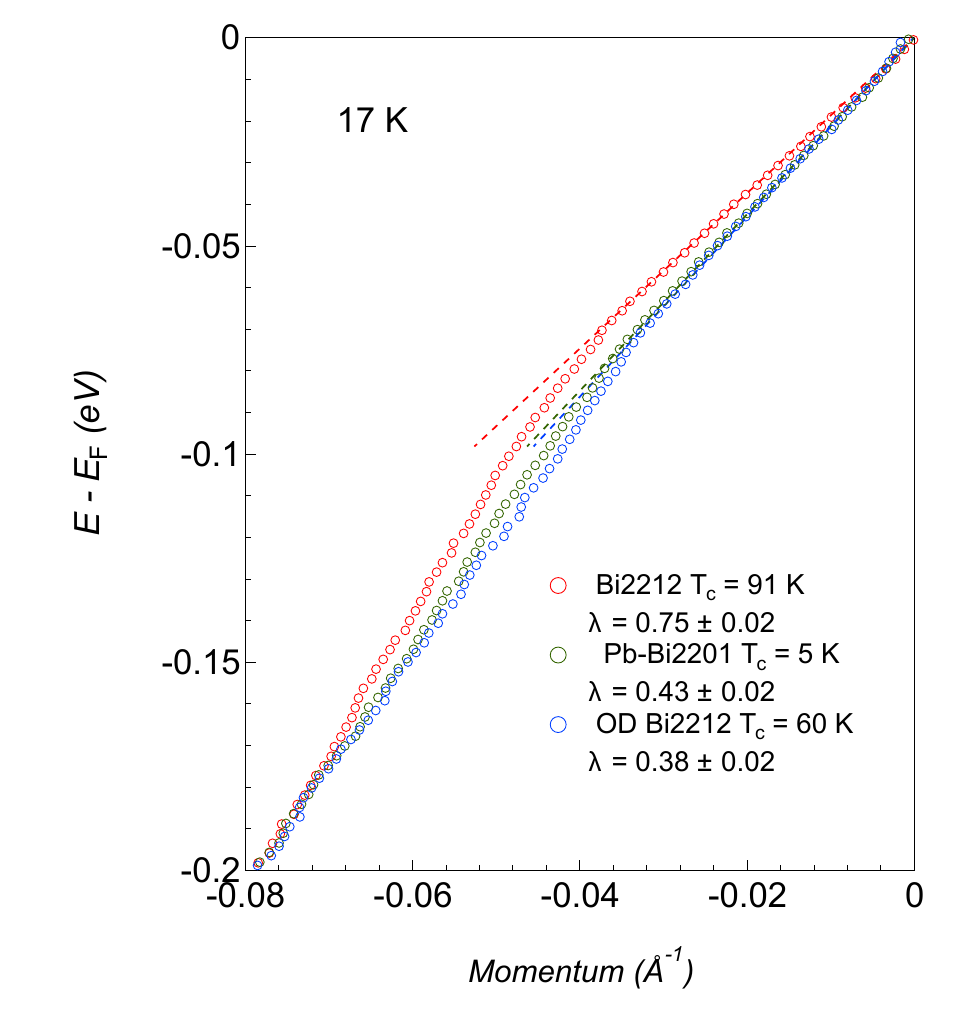}
  \caption{
   Dispersion. Nodal dispersions of an optimally doped Bi2212 ($T_\text{c}$ = 91 K), a heavily overdoped Bi2201 ($T_\text{c}$ = 5 K), and a heavily overdoped Bi2212 ($T_\text{c}$ = 60 K) measured at 17 K. Dispersions at the Fermi level and at the binding energy -0.2 eV are scaled together to compare their kink strength. Experimental coupling strength λ is related to the ratio of velocities of the bare $v_0$ and the dressed ($v_\text{F}$) electrons by $\lambda=v_0/v_\text{f}-1$.}
\end{figure}

\begin{figure}\centering\includegraphics{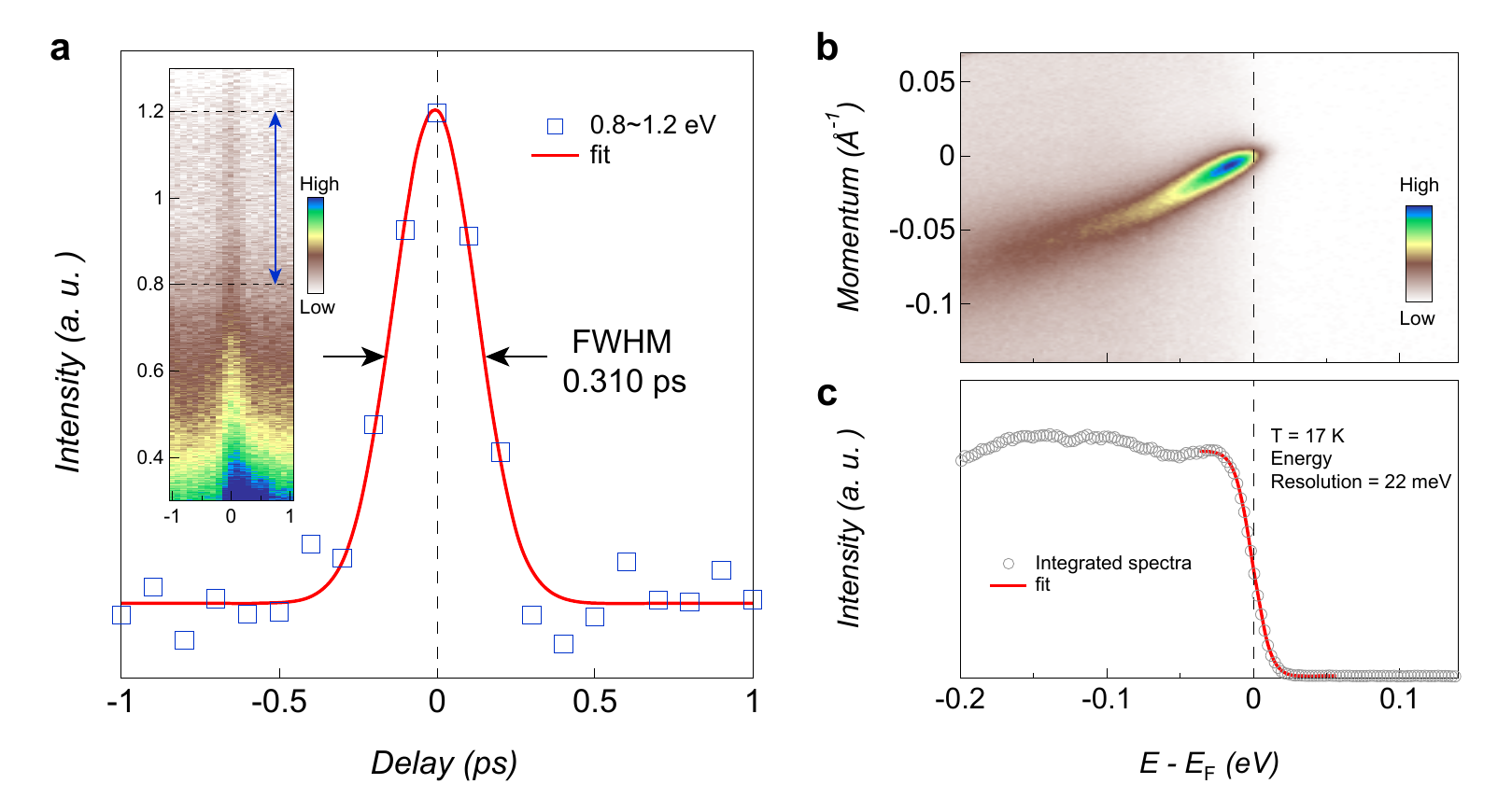}
  \caption{
   Time resolution and energy resolution. \textbf{a}, The energy-integrated (0.8 - 1.2 eV) pump-probe spectr\textbf{a}, measured on polycrystalline gold at room temperature with a pump fluence ~35 $\mu$Jcm$^{-2}$, versus the delay time between the pump and probe pulses. The red solid curve is a fit to a Gaussian function, giving a time resolution of ~310 fs. Inset is the related to the Fermi energy. \textbf{b}, Nodal equilibrium spectra of an optimally doped Bi2212 measured at 17 K. \textbf{c}, Spectra in b integrated through the momentum axis. The red solid curve is a fit to an energy-resolution-convolved Fermi-Dirac distribution function at 17 K. The obtained energy resolution from fitting is ~22 meV.}
\end{figure}